\documentclass[reprint,amssymb, amsmath, aip,cha,twocolumn]{revtex4-1}
\usepackage{graphicx}
\usepackage{color}
\usepackage{epsfig}
\usepackage{bm}%
\usepackage[colorlinks=true,linkcolor=blue]{hyperref}%
\expandafter\ifx\csname package@font\endcsname\relax\else
 \expandafter\expandafter
 \expandafter\usepackage
 \expandafter\expandafter
 \expandafter{\csname package@font\endcsname}%
\fi
\begin{document}
\title{Melting transition of two-dimensional complex plasma crystal in the DC glow discharge}%

\author{S. Jaiswal }
\email{surabhijaiswal@gmail.com}
\affiliation{Physics Department, Auburn University, Auburn, Alabama 36849, USA.}%
\author{Ed Thomas Jr.}
\affiliation{Physics Department, Auburn University, Auburn, Alabama 36849, USA.}%

\date{\today}
\begin{abstract}
The formation of self-consistent dust crystal and its melting is a well known phenomenon in rf generated plasma but remains challenging in DC glow discharge plasma. Here, we report the melting of a two dimensional dusty plasma crystal, suspended in the cathode sheath of a DC glow discharge plasma. The experiments are carried out in a $\Pi-$shaped Dusty Plasma Experimental (DPEx) device where a stationary crystal of melamine formaldehyde particles is formed between the confining strips in a background of Argon plasma. The stable structure breaks and leads to a fluid state on reducing the neutral pressure.
The neutral pressure range where this melting transition is observed is an order of magnitude less than what reported in rf discharge plasma. The transition is confirmed by evaluating the variation in different characteristic parameters such as the pair correlation function, voronoi diagram, local bond order, defect fraction and dust temperature as a function of background neutral pressure. The transition is attributed to an increase in effective particle temperature which we believe is occurred due to increase in charge fluctuation and ion streaming. The special feature of the device that helps formation of dust crystal and its melting in DC glow discharge can be implemented to study various phenomena associated with dust crystal in dc glow discharge plasma devices.    
\end{abstract}
\maketitle
\lq\lq Dusty" or \lq\lq complex" plasmas composed of the usual combination of electrons, ions, neutral atoms and the charged, dust particulates of micron or submicron size. These particles, which either added into the plasma or produced by sputtering or by polymerization and agglomeration of reactive gases, get negatively charged by collecting more electron than ions  and act as a third component adding more richness to the collective dynamics of the system. Their existence in astrophysical situations such such as planetary rings, comet tails, interplanetary media, and interstellar clouds \cite{goertz, nakamura_book} as well as various industrial application \cite{selwyn} make them an active area of research for the last few decades \cite{melzer_book, morfill}. The highly charged dust cloud in the laboratory plasma levitates near the sheath boundary by balancing the electrostatic force due to the sheath electric field and the gravity. They interact by means of their Coulomb repulsion and therefore form a strongly coupled system. The strength of coupling can be determined by a coupling parameter $\Gamma$, which is the ratio of the interparticle coulomb potential energy to the dust thermal energy. When $\Gamma$ exceeds a critical value, the system formed a regular lattice structure so-called coulomb crystal. 
The necessary plasma condition for the dust crystal formation was first derived by Ikezi \textit{et.al}\cite{ikezi1986} and the  experimental observation was first reported in 1994 by Chu \textit{et.al}\cite{chu_1994} and Thomas \textit{et.al}\cite{thomas1994}. These observation opened up a whole new area of research. The control over coupling strength of the system made dusty plasma an excellent platform for studying the phenomenon of phase transition, transport properties of the strongly coupled system and other related topic that has great applicability in variety of areas such as statistical
mechanics, soft condensed matter and colloidal suspensions. 
The spatial and time scales of the particle motion allow them to be visualized using laser illumination and high-speed cameras. Further, the weak frictional damping enables the measurement of the dynamics and kinetics of individual particles. Such a diagnostic convenience of dusty plasma motivated a great deal of research related to crystal formation, phase transitions of two and three dimensional crystals\cite{melzer_crystal_1996, schweigert1, schweigert2,numkin}, re-crystallization\cite{christina}, instabilities\cite{lenaic, meyer}, magnetic field effect on the phase transition\cite{jaiswal_magnetic}, heat transport\cite{nonumura_heat, fortov_heat} and many more. \par
Mostly, the experimental studies have been devoted to the formation of the dust crystal and its melting dynamics is done using rf discharge plasma. The studies on dust crystal in DC glow discharge plasma is very limited. Among such few works are those of Fortov \textit{et al.}\cite{fortov_dc_crystal} and Maiorov \textit{et al.}\cite{maiorov}. In both the experiments, a crystalline structure with multiple layers were formed by the trapping of dust particles in the standing strata of a dc glow discharge that appear under weakly ionized conditions in the positive column of a DC glow discharge. But this method has the
 limitations that the crystal structure can be formed and sustained over a narrow range of discharge parameter
and the strong variation in the electric field lead to small scale inhomogeneities in the particle cloud. Mitic \textit{et al.}\cite{mitic_dc_crystal} studied the formation of three dimensional ordered structure in the low frequency DC plasma at the discharge condition where no striation were present. Particle cloud in their experiment were trapped by mitigating the effect of longitudinal electric field by using an alternating DC current of frequency of 1 kHz. These experimental arrangements were still found to be insufficient for the observation of a perfect crystal structure in DC glow discharge plasma. Reportedly, there are two main factors viz. lower charging of the dust particles and exorbitant heating of the dust by ion bombardment makes it difficult to formed a perfect crystalline structure in DC plasma. Due to insignificant electron density in the sheath of the DC plasma the particle which immersed deep inside the sheath cannot collect many electrons \cite{melzer_book}. Therefore, the particle can be levitated only near the sheath edge and the charged acquired by the particle are considerably less to those compared to one in the rf sheath which collapses and expands at the applied radio frequency so that the time-averaged electron density is everywhere non-zero in the sheath and hence the particle maintain a negative charge at every location of the sheath. This way the coupling parameter for the same dust temperature tends to be relatively lower in the case of DC plasma. The situation of heating of the dust due to the impact of ion streaming is more effective in the case of parallel electrode arrangement because in this case the ions accelerated by the DC voltage  directly impinge on the levitated dust cloud that is formed between the electrodes and causes its temperature to rise. Therefore, any such improvement in the geometry of the electrode so that the direct bombardment of ions on the dust cloud can be restricted, could overcome the difficulty of the formation of dust crystals in such devices. 
Recently, the formation of a stable plasma crystal in the DC glow discharge has been observed \cite{surabhi_thesis, hari_pop_2018} in the dusty plasma experimental (DPEx) device where an asymmetric electrode arrangement with extra confinement strip/ring is used for particle confinement \cite{jaiswal2015}. A phenomenon of of melting transition of dust crystal in DC glow discharge is still unexplored. The detailed investigation of crystalline behavior and phase transition of dusty plasma in the DC discharge is important for the point of understanding of fundamental physics as well as for production and control of the properties of the dust crystal in different plasma sources.\par
In this paper we report the first experimental observation of melting transition of two- dimensional dusty plasma Coulomb crystal levitating in the cathode sheath of the DC glow discharge plasma. The experiments have been carried out in the dusty plasma experimental (DPEx) device \cite{jaiswal2015} with a unique electrode configuration that permits investigation of dust Coulomb crystals over a wide range of discharge parameters with excellent particle confinement. The melting is induced by changing the discharge parameter. The phase transition is confirmed by investigating the global and local variation in the structural dynamics of the crystalline structure by calculating the pair correlation function, local bond order parameter and defect fraction. The cause of melting is qualitatively explained by increasing temperature as a combined effect of ion streaming and charge fluctuation.
\par
The paper is organised as follows: in the next section, in Sec.~\ref{expt}, we describe the experimental arrangement in detail. In Sec. \ref{result}, we discuss the experimental results on the formation of the dust crystal and its melting dynamics. A brief concluding remark is made in Sec.~\ref{conclusion}.
\section{Experimental arrangement\label{expt}}
\begin{figure}[ht]
\centering{\includegraphics[scale=0.35]{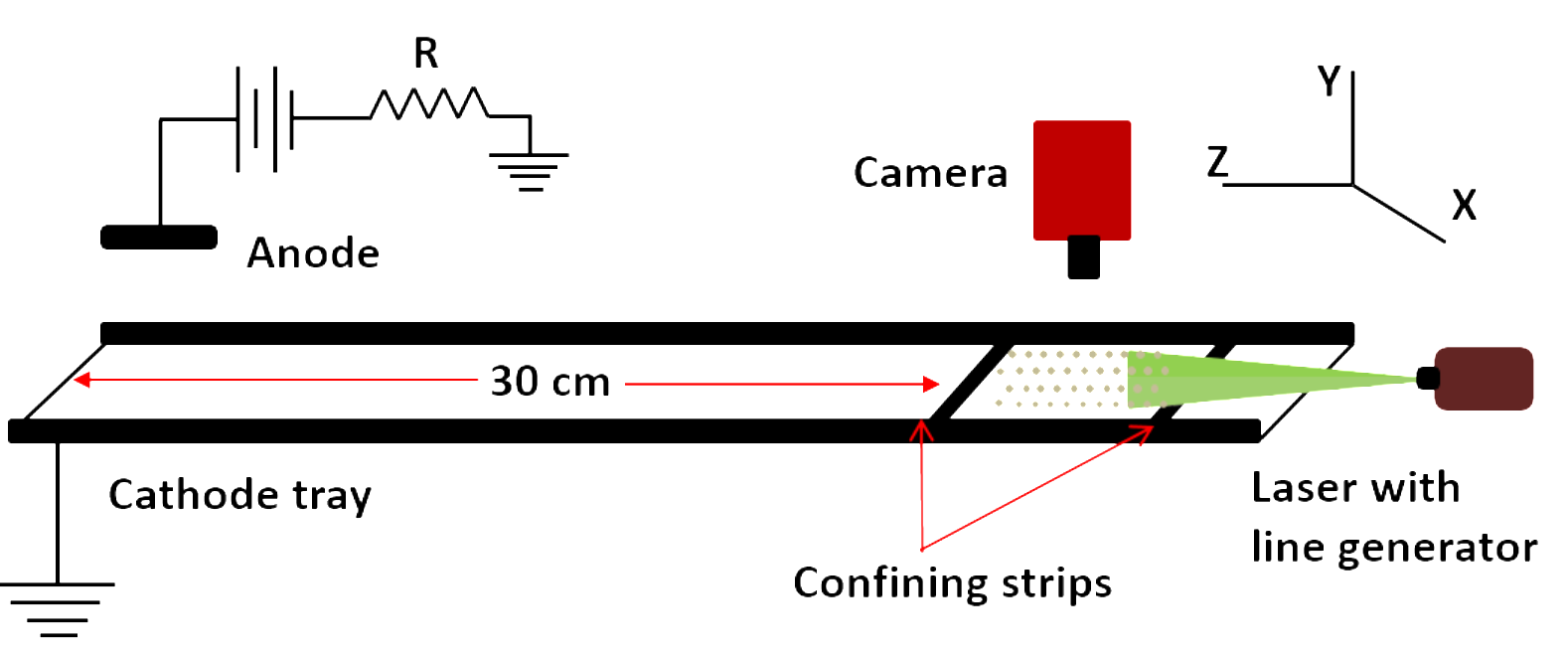}}
\caption{(a) Schematic diagram of Dusty Plasma Experimental (DPEx) setup., R: current limiting resistance. }
\label{fig:fig1} 
\end{figure}

The experiments are performed in Dusty Plasma Experimental (DPEx) device which consists of {$\Pi-$} shaped pyrex glass tube with a horizontal section of 8 cm inner diameter and 65 cm length. This section is used for all the dusty plasma experiments. The two vertical section of same diameter and 30 cm length is used for other functional access such as gas inlet, pumping connection etc. An asymmetric electrode arrangement, consist of a 3 cm stainless steel disc and 40 cm $\times$ 6.1 cm $\times$ 0.2 cm cathode tray is used to strike argon plasma. Fig.~\ref{fig:fig1} shows the schematic diagram of the experimental arrangement used for current experiment. A detailed description of the experimental setup and its operational characteristic is reported elsewhere (see ref. \cite{jaiswal2015}). \par

The experimental setup is evacuated to a base pressure of $10^{-3~}$ mbar by a rotary pump. To remove any kind of impurity, the Argon gas is then flushed several times and pumped down to base pressure. Finally, the working pressure is set in the range of $0.1-0.2$~mbar by maintaining the pumping speed and the gas flow rate. A DC glow discharge plasma is formed in between the anode and cathode by applying a discharge voltage in the range $310-320$~volt. The discharge current is measured as 2-5 mA. The mono-disperse melamine-formaldehyde spheres of diameter $4.38\pm 0.06~\mu$m are introduced into the plasma by shaking the dust dispenser which is inside the vacuum vessel. These particles get negatively charged by collecting more electrons than ions and get trapped in the plasma sheath boundary above the grounded cathode. In this levitating condition, the vertical component of the sheath electric field provides the necessary electrostatic force to the particles to balance the gravitational force. The radial sheath electric fields due to the bending edges of the cathode tray are responsible for the radial confinement of the dust particles against their mutual coulomb repulsive forces. 
Two stainless steel strips of 1 cm$\times$ 6 cm, placed at 6 cm apart on the cathode, facilitates the axial confinement through axial sheath electric field. In this way a steady-state equilibrium dust cloud is levitated over the cathode in between the two stainless steel strips at a height of $~\approx$ 2 cm. The levitation height and cloud size can be modified by varying the discharge parameter and hence manipulating the sheath around the rectangular region. The axial confinement also influences the dynamics of the streaming ions in a significant manner. A detailed description of such a behavior in DPEx is provided in the study by Hari \textit{et al.}\cite{hari_pop_2018}. The particle-cloud is illuminated by a horizontally expanded thin sheet of green laser light (532 nm, 100 mW) which is sufficiently constricted vertically to study individual layer of the dust-cloud. The Mie-scattered light from the dust particles is captured by a CCD camera (shown in Fig~\ref{fig:fig1}) at 25 fps with a resolution of 9 $\mu$m/pixel and the images are stored into a high speed computer.
%
\section{Experimental Resutls and discussion \label{result}}
In rf plasma the melting transition of the dust crystal depends mainly on two parameters such as rf power and background pressure but also depends on the number of layers formed. Larger number of layers corresponds to more disordered crystalline structure. The layer which is closer to the electrode melts first and shows more randomized particle motion whereas the uppermost layer remains in a specific state for a longer period of time. In DC we found that the transition depends on discharge voltage and background pressure. It has been observed that the small variation in discharge voltage changes the plasma condition abruptly therefore a clear transition from solid to liquid in a sequential order is difficult to observe. Therefore, we fixed the voltage at 310 volt and perform the experiment by varying the pressure. It is worth mentioning that we have also seen two layers in the glow discharge plasma however, most of the crystal points do not show any deviation; this indicates that particles are almost aligned in the vertical direction also and lower layer does not make any significant effect on the upper layer. The structure is almost stationary with a small thermal fluctuation around the equilibrium position that we have checked from the overlapped image of 100 consecutive frames. In our experiment we mainly focused on the top most layer.\par
\begin{figure}[!htbp]
\centering{\includegraphics[scale=0.3]{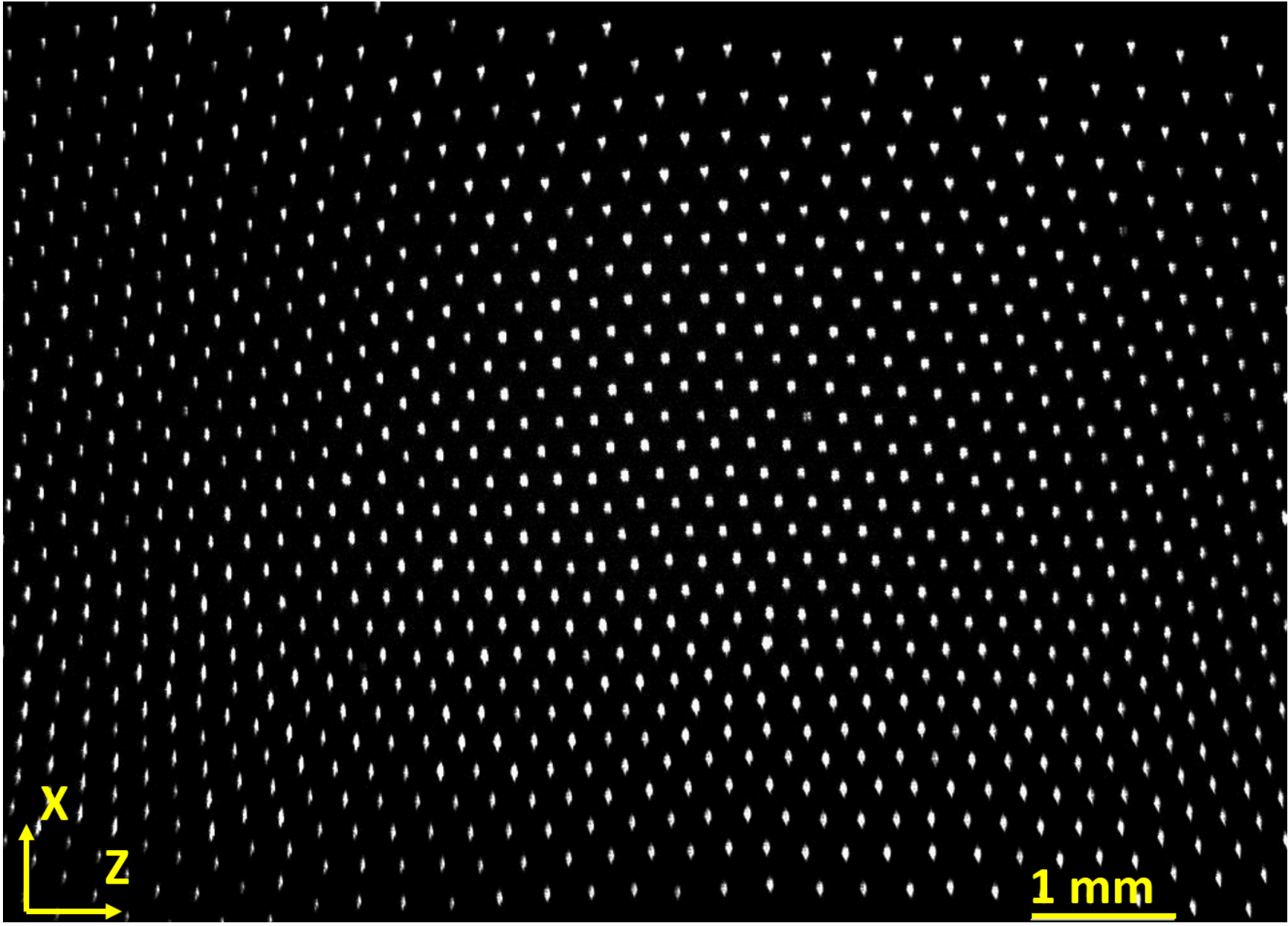}}
\caption{ A snapshot of the coulomb crystal formed at pressure 14 pascal, and discharge voltage of 310 volt.}
\label{fig:fig2} 
\end{figure}
\begin{figure}[!htbp]
\centering{\includegraphics[scale=0.3]{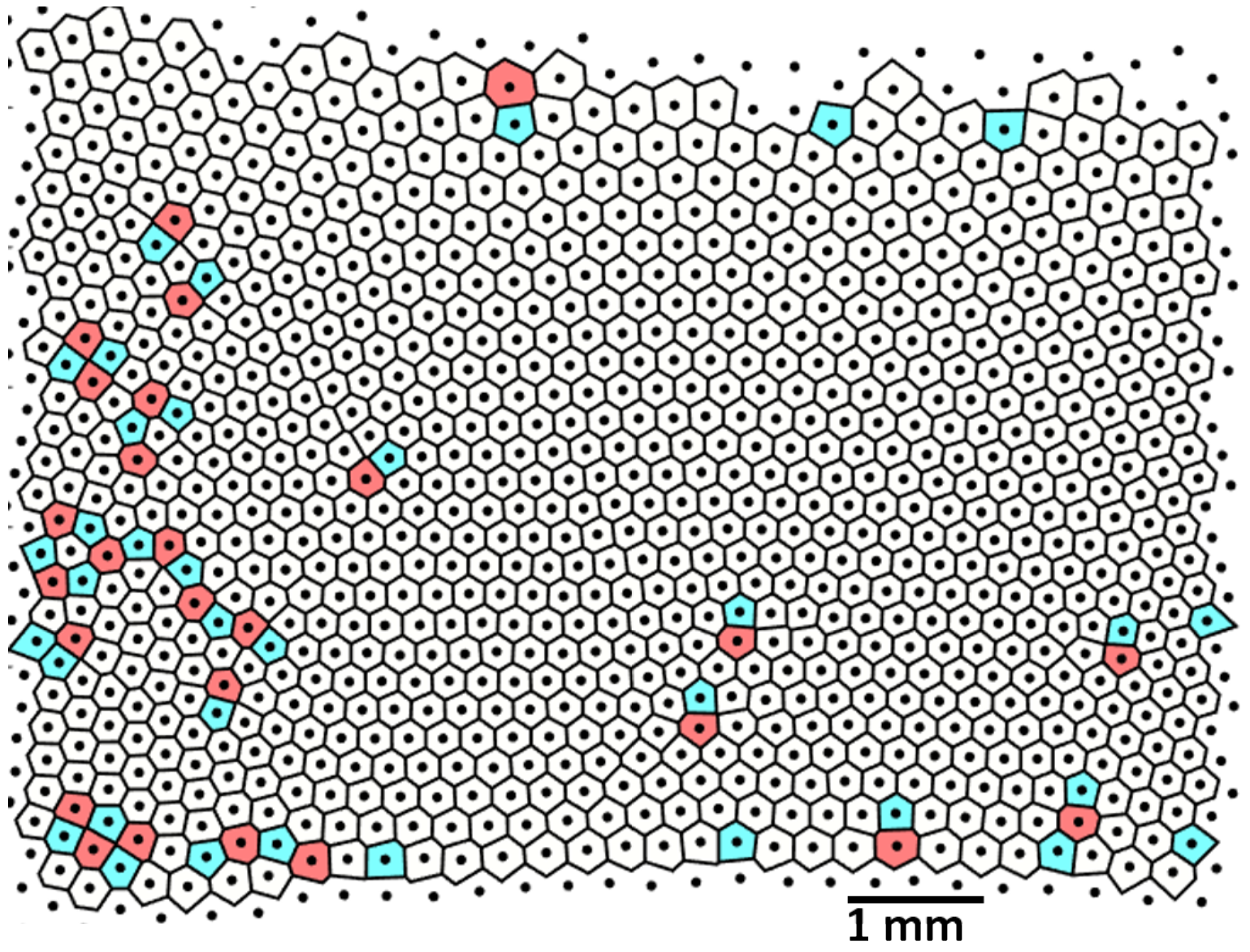}}
\caption{ Voronoi diagram of the particle location at p = 14 pascal which is shown in Fig~\ref{fig:fig2}.}
\label{fig:fig3} 
\end{figure}
\begin{figure}[!htbp]
\centering{\includegraphics[scale=0.45]{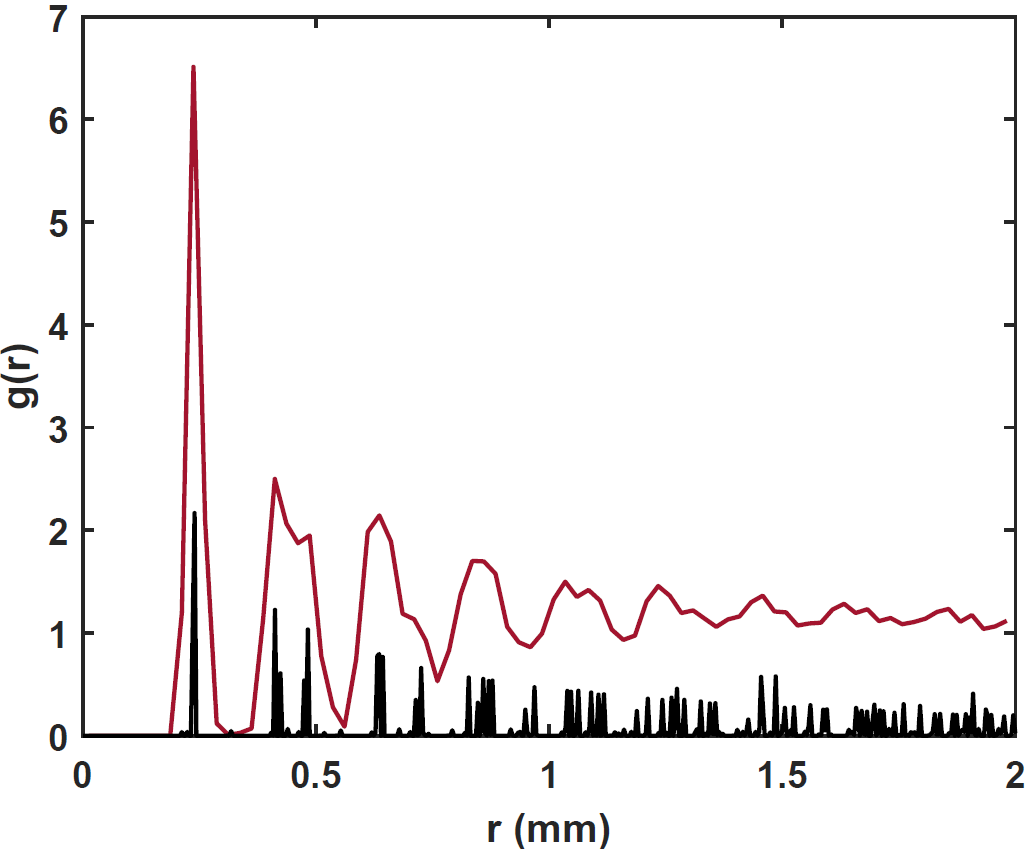}}
\caption{ Pair correlation function g(d) of the particle cloud at 14 pascal. The discharge voltage is 310 volt. The black vertical lines showing the positions for an ideal Hexagonal lattice.}
\label{fig:fig4} 
\end{figure}

By careful variation in pressure while keeping the voltage fixed we found that the dust cloud settled into crystalline structure at a pressure of 14 pascal. A snapshot of coulomb crystal is shown in Fig~\ref{fig:fig2}. It can be clearly visible that dust particle arranged into a regular hexagonal lattice with nearly uniform particle spacing and good transnational periodicity which represents that system has long range ordering. The voronoi diagram (Wigner-Seitz cell) of Fig~\ref{fig:fig2} is shown in Fig~\ref{fig:fig3}. From this analysis we can quantify the hexagonal structure and can obtain a rough estimate of ordering of the crystalline structure. It is seen that most of the cells are six sided with almost all having the same size. Some of the polygons in the structure have five and seven sides as shown by red and cyan color in the Fig~\ref{fig:fig3} 
which represents the disclination in the crystal. An array of disclinations are visible mostly in the left region of the structure which can leads to  defects in the crystal as previously reported in the crystal formed in rf plasma \cite{melzer_crystal_1996}. In the diagram, it is calculated that the 93.2$\%$ of the Wigner-Seitz cells are hexagon whereas $3.9\%$ and $2.9\%$ polygon have 5 sides and seven sides respectively. It indicates the observed structure is a highly ordered crystal with very few defect. The percentage of ordering can be calculated by dividing the number of hexagon to the total number of polygons in the Voronoi diagram. The structural order parameter is estimated as 93$\%$. We can further estimate the range of positional ordering of the crystalline structure by calculating the radial pair correlation function, g(r). It is a measure of probability of finding the particles at a distance r from the reference particle and can be calculated by directly measuring the average distance between particles. Fig~\ref{fig:fig4} depicts the g(r) vs r of the structure shown in Fig~\ref{fig:fig2} where 50 frames have been used for averaging. It can be seen from the figure that the positional order of the structure is preserved upto 6 nearest neighbor which reveals a long range ordering and hence a perfect crystal formation at 14 pascal. The black vertical lines in the bottom showing the position of an ideal hexagonal lattice which is in line with corresponding g(r) of the observed crystal. Therefore observed structure is a hexagonal structure also resembles to those reported in rf plasma by Melzer \textit{et. al.}\cite{melzer_crystal_1996}. The pair correlation function also provides a good estimate for the mean inter-particle spacing from the position of the first peak. The mean inter-particle distance for the structure in its crystalline states comes out to be $\sim$ 250 micron.
\subsection{Structural change of the cloud with changing pressure  \label{single}}
\begin{figure}[!ht]
\centering{\includegraphics[scale=0.6]{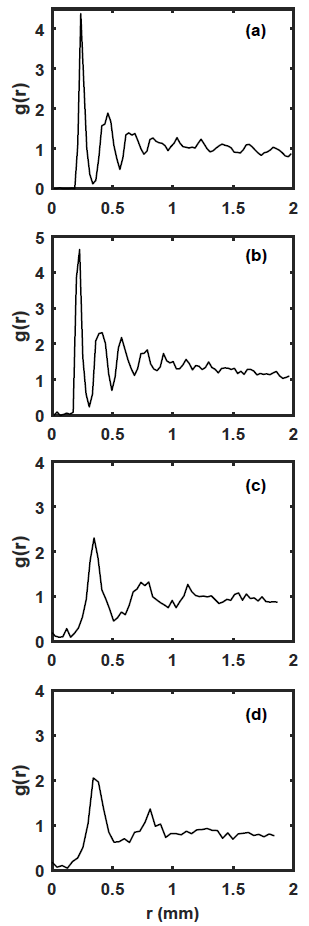}}
\caption{ The pair correlation function g(r) of the particle clouds with the
changing pressure (a) 14 Pa, (b) 13 Pa, (c) 12 Pa, and (d) 11 Pa.}
\label{fig:fig5} 
\end{figure}

We now analyze change in the structure of the particle cloud with lowering pressure. 
In order to get a better understanding of melting transition of the dust crystal in DC plasma, we performed variety of diagnostic test on the experimental data in the form of calculating the pair correlation function,  defect measurement, calculation of local bond order ($\Psi_6$) and estimating the dust temperature. These parameters gives the information of global as well local structural properties of the cloud. Below we provide a detailed description of the results of our analysis. We kept the discharge voltage constant at 310 volt while reducing the background neutral pressure from 15 to 11 pascal in the unit of 1 pascal. 
The pair correlation function which, delineate the global system properties in terms of degree of long range order of the distribution of particles, is plotted in Fig~\ref{fig:fig5}. Figure ~\ref{fig:fig5}(a) presents the g(r) vs interparticle distance (r) at pressure of 14 pascal. The nature of the correlation function shows the existence of long range ordering over a distance upto 2 mm with a very pronounced peak which is indicative of the system being in the crystalline state.  
This can also be confirmed from the splitting of the second peak. With the decrease in pressure, the peaks become shorter and flattered, showing the increasing disorderness of the cloud, and also the number of peaks or the length upto which ordering is observable gets decreased. Fig~\ref{fig:fig5}(c) shows the pair correlation function at P = 12 Pa and illustrates that the phase state of the particle changes significantly and it tends towards the liquid state with the primary peak followed by a fast descending second or third peak. At a pressure of 11 Pascal, the system becomes almost a liquid like that can be clearly seen in Fig~\ref{fig:fig5}(d), where only a small hump appears after a primary peak in the correlation function. So these pair correlation function indicate a transition from ordered solid structure to the fluid at lowering pressure. It is quite interesting to note that 
 the stability of the crystalline structure in DC plasma maintained only upto a certain range of background neutral pressure and a transition to disordered state occurs at small variation of discharge parameters 
 unlike the rf plasma where a stable crystal can be sustained upto a long range of pressure variation. \par
\begin{figure}[!htbp]
\centering{\includegraphics[scale=0.5]{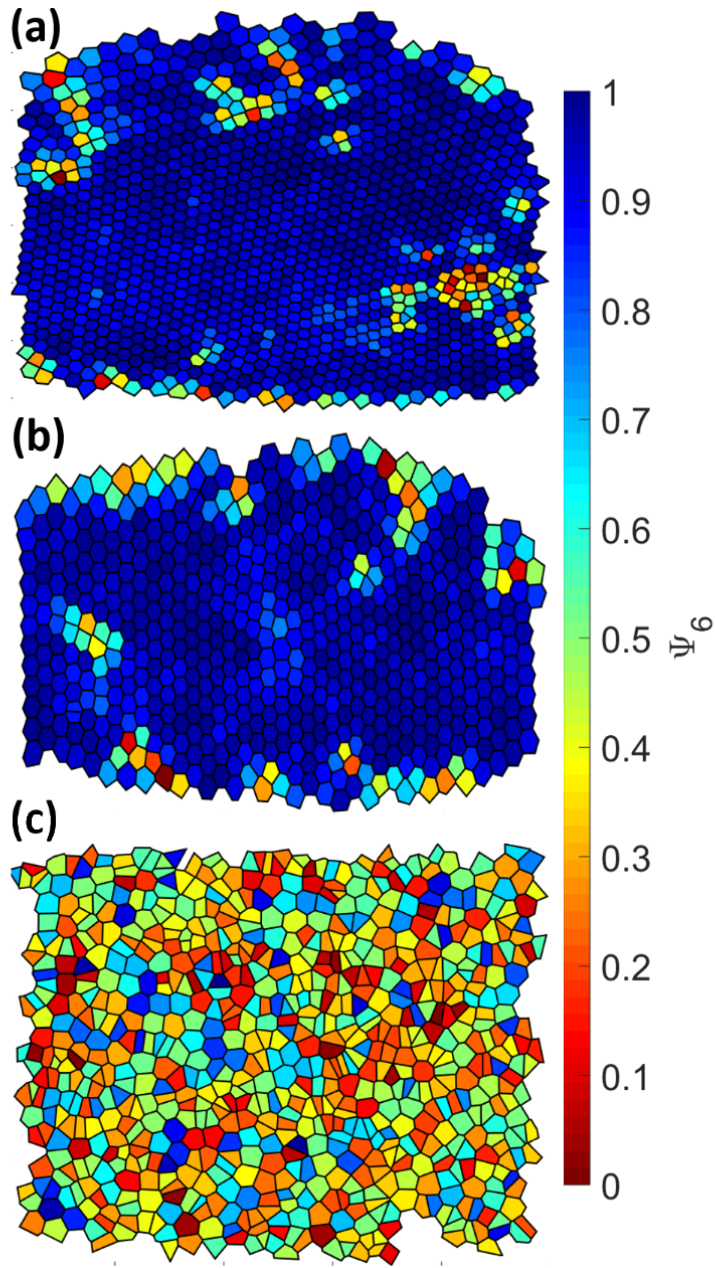}}
\caption{ Development of the local order parameter $\Psi_6$ during melting transition. (a) corresponds to 15 pascal. (b) 13 pascal and (c) 11 pascal. Unit cells of particles are colored according to their bond-order parameter.}
\label{fig:fig6} 
\end{figure}

\begin{figure}[!htbp]
\centering{\includegraphics[scale=0.4]{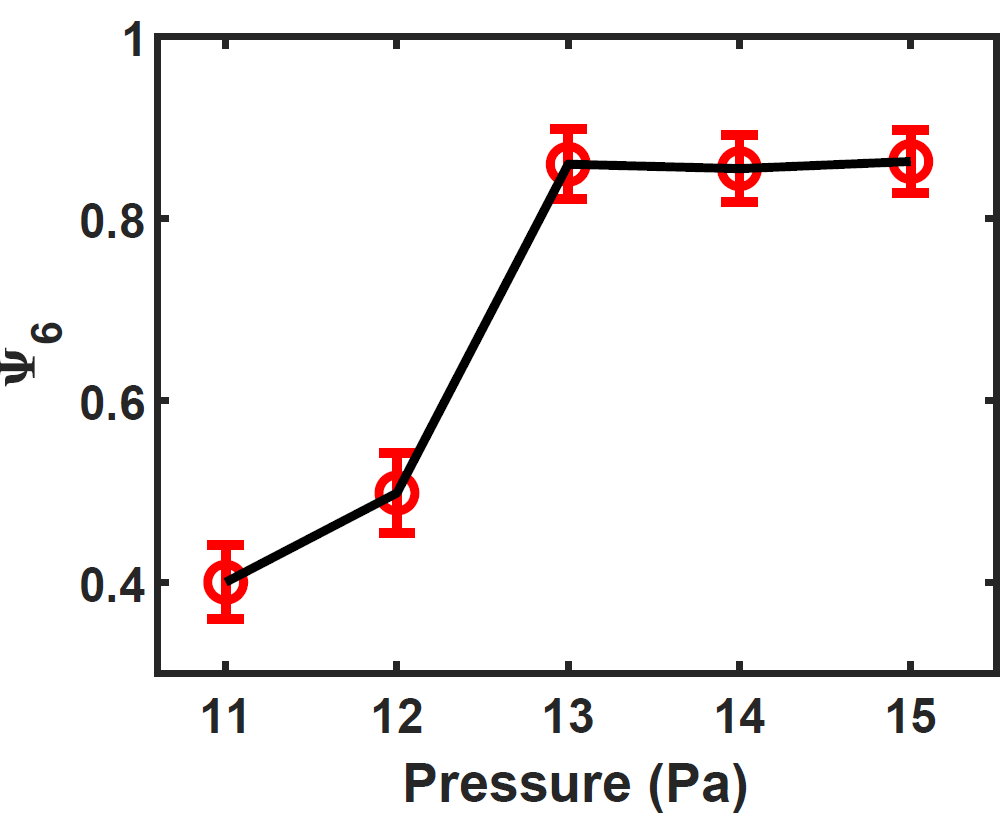}}
\caption{ Variation of average bond order parameter $\Psi_6$ with pressure. }
\label{fig:fig7} 
\end{figure}

In order to get a deep insight on melting dynamics of the structure it is important to understand the variation in structural dynamics of the particle cloud locally. To do that we have calculated the local bond order parameter and the average defect fraction. These quantities are defined locally and measure the structural properties of the lattice directly at the respective particle positions \cite{christina_2007, christina_thesis}. A bond order parameter, $\Psi_6$ examine the lattice in terms of the local orientational order of particles \cite{christina_thesis, coudel_2018}. For a given particle k, $\Psi_6$ is defined as 
\begin{equation}
\Psi_6(k) = \frac{1}{N}\left|\sum_{n=1}^{N}e^{6i\theta_{kn}}\right|,
\end{equation}
where N is the number of nearest neighbors and $\theta_{kn}$ is the
angle of the bond between the kth and nth particles with respect to the x-axis. For an ideal hexagonal structure the bond order parameter is maximum ($\Psi_6 = 1$) hence it is a good measure of the deformation of a cell from the perfect hexagon. Lattice sites with nearest neighbor bond angles deviating from 60$^\circ$ will decrease $\Psi_6$. The same accounts for lattice sites with
a number of nearest neighbors other than six. For calculating the bond order parameter, we first identify the particle locations and find out the nearest neighbors of all the particles using Delaunay triangulation. After that for every particle k we calculate a vector going to its nearest neighbours n and then calculated the angle of the vector kn with respect to the x-asis. A voronoi map of the particle cloud which is color coded according to $\Psi_6$ is shown in Fig~\ref{fig:fig6}. Fig~\ref{fig:fig6}(a) presents the voronoi map corresponding to the pressure of 15 pascal. As we can see from the figure that the structure shows a high orientation ordering as most of the voronoi cell around the particles exhibits perfect hexagonal structure (colored by blue). Those voronoi cells which exhibit very low ordering, as visible in the right side of the cloud, cause defects. The number of cell having lower bond order is seen to be increased as we decrease the pressure to 13 pascal (Fig~\ref{fig:fig6}(b)) however the change is not very significant. At a pressure of 11 pascal the cloud looses its orientational ordering completely and almost all the cells manifest a lower bond order as can be seen in Fig~\ref{fig:fig6}(c).
 The variation in average bond order parameter, $\Psi_6$ with pressure is plotted in Fig~\ref{fig:fig7}. The mean $\Psi_6$ which is obtained by calculating the $\Psi_6(k)$ for each unit cell and then taking the average 
 over all the cell gives the information of overall orientational ordering of the system. The local order of the particle cloud is  $\sim$ 0.87 at 15 pascal which is almost constant upto 13 pascal. It decreases to 0.48 when we reduce the pressure to 12 pascal. The minimum value of $\Psi_6$ is comes out to be 0.4 for 11 pascal which is close to the values reported for the melting dynamics in rf plasmas.\cite{christina_thesis}\par
We further parameterize the structure by calculating the defect. Generally, a defect causes the disorganization of particles in a crystal lattice. Most common defect in two dimensional system are point defects or disclinations which arises due to vacancy or an interstitial.
 A hexagonal lattice with typically six nearest neighbors to each lattice site predominantly encounter two types of point defects namely \lq five-folded' or \lq seven folded' lattice sites. Other kind of defects exist in a system are \lq\lq free dislocation", which is an additional row of dust particles in an ideal lattice that can be viewed as a pair of a 5-fold and 7-fold disclination, and dislocation pairs (quarter of 5-fold and 7-fold disclination). A variety of theories has been hitherto reported for rf plasma that describe the melting transition of dust crystal based on the formation of these defects \cite{melzer_crystal_1996, schweigert1, christina_thesis}. The KTHNY theory \cite{melzer_crystal_1996, christina_thesis} describe the melting of the crystal lattice by two continuous phase transition in which firstly the dislocation pairs unbind into free dislocations followed by the stage where dislocations (pair of a 5-fold and 7-fold disclination) breaks into free disclinations. 
 The first stage is called as hexatic phase where positional order is destroyed but orientation order still can be found in the system. In the second stage the orientation order also disturbed and the liquid state is reached. In our case, we have performed Delauney triangulation to find out the position of each particle and locate the nearest neighbor of corresponding particles. We then find out the position and count the particle having five and seven nearest neighbors respectively. The defect fraction can be calculated by taking the ratio of number of disclinations (5 or 7 folds ) to the total number of particles in a frame.
\begin{figure}[!htbp]
\centering{\includegraphics[scale=0.4]{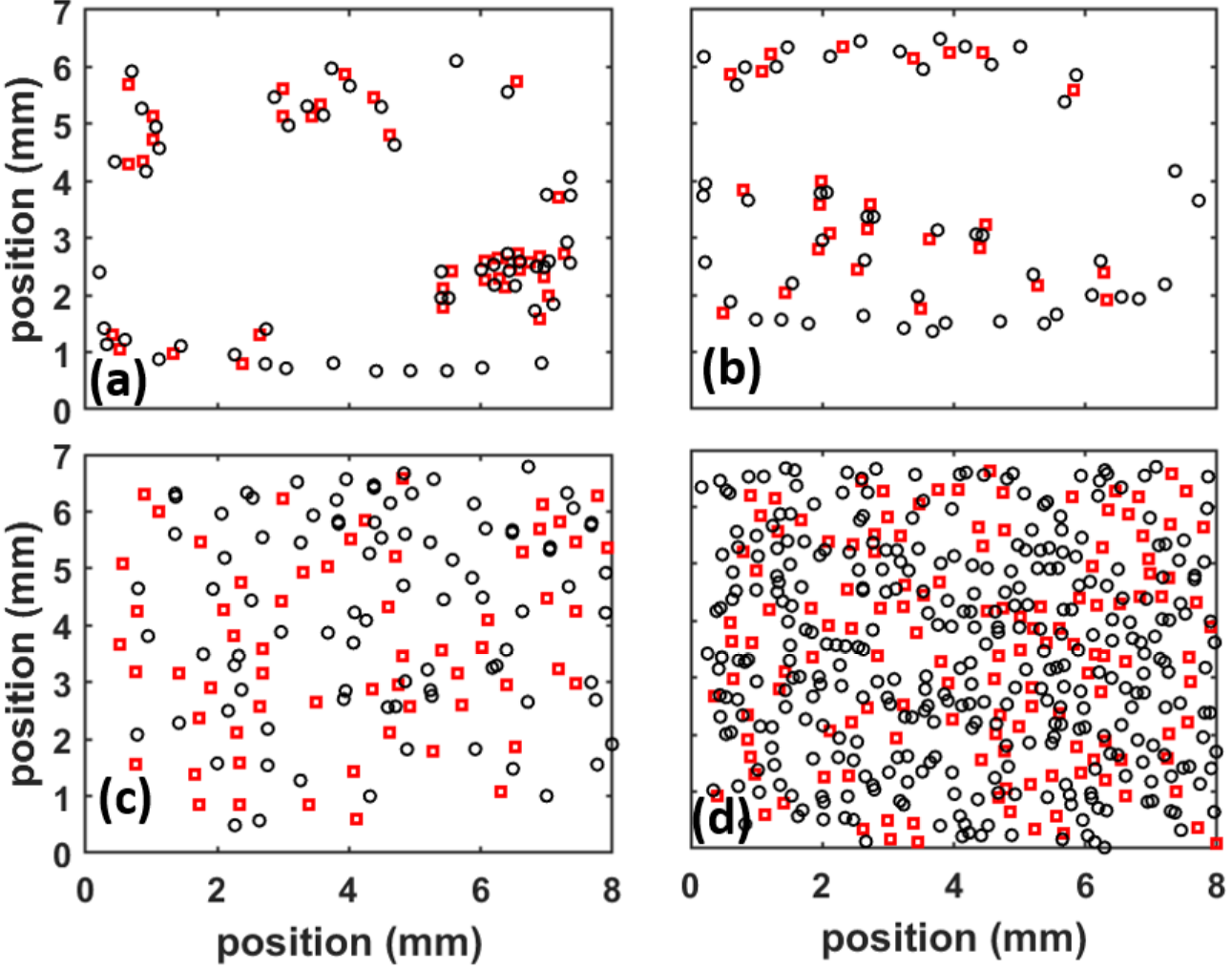}}
\caption{ Defect plots for the particle cloud at (a) 15 Pascal, (b) 13 Pascal (c) 12 pascal and (d) 11 Pascal. The discharge voltage is 310 volt. Square and circle denotes the seven sided polygon and pentagon respectively.}
\label{fig:fig8} 
\end{figure}
Fig~\ref{fig:fig8} shows the defect plots at various pressures. The red circle 
and black square in the figures mark the locations of 5 and 7-fold defects.  At pressure of 15 pascal (see Fig~\ref{fig:fig8}(a)) we can see the dislocation pairs but mostly free dislocations can be seen. Some chainlike arrangements of the defects also appear. The defect fraction corresponding to five and seven nearest neighbors are calculated as 4.2$\%$ and 2.9$\%$ respectively. There is no significant variation found upto 13 pascal with the defect fractions of 6.3$\%$ (five nearest neigbours) and 3.2 $\%$ (seven nearest neigbours). A notable change in the defect fraction is observed as the pressure is further reduced to 12 Pascal and the defect arrange more and more in chains as visible from Fig~\ref{fig:fig8}(c). The defect fraction in the case is calculated as 30.6$\%$ and 20.6$\%$ respectively. At 11 Pascal almost all the defect sites corresponds to free disclination (Fig~\ref{fig:fig8}(d)) and defect fraction for 5 fold is calculated as 38$\%$ whereas 29$\%$ for 7- fold. The observed melting transition does not exactly follow the KTHNY theory of melting.\par

\begin{figure}[!htbp]
\centering{\includegraphics[scale=0.40]{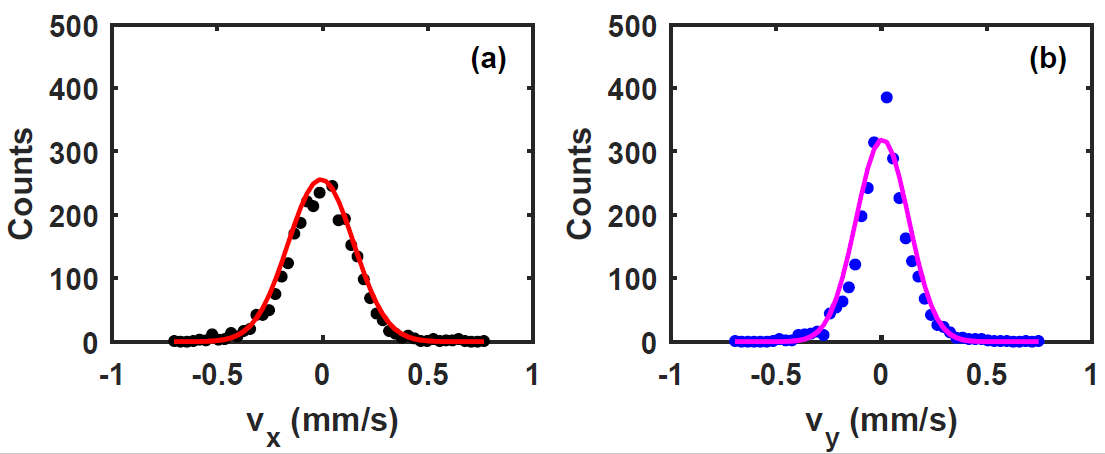}}
\caption{A typical distribution of velocities is depicted for (a) the x component, (b) the y component. The points depict the distribution of measured velocities, while the solid curve represents the fit of a Maxwellian velocity distribution.}
\label{fig:fig9} 
\end{figure}
\begin{figure}[!htbp]
\centering{\includegraphics[scale=0.4]{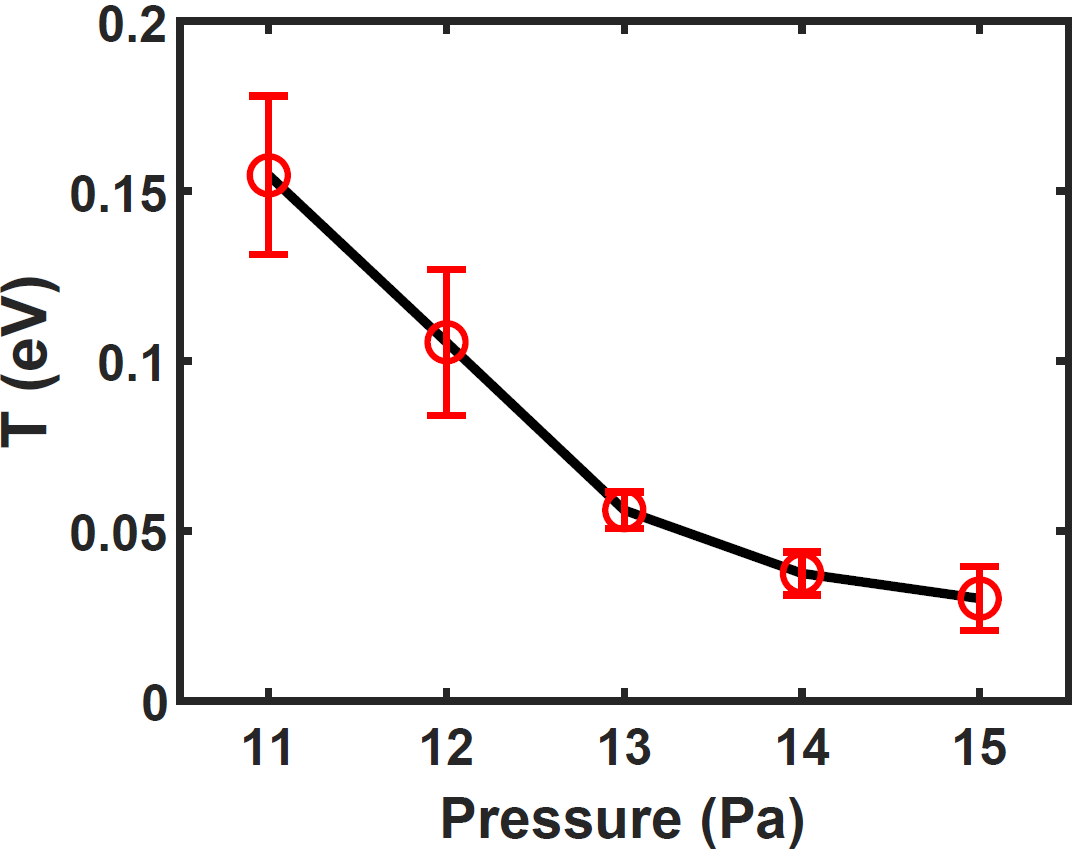}}
\caption{ Plot of the kinetic temperature of the dust cloud as a function of the neutral gas pressure}
\label{fig:fig10} 
\end{figure}

For the experimental measurement of the kinetic temperature a velocity distribution function in two orthogonal direction has been calculated. For a better statistics, a video sequence of 100 frames has been chosen where a particle tracking velocimetry (PTV) analysis package, DAVIS 8 \cite{davis} is used to construct the velocity vector fields. Fig~\ref{fig:fig9} shows the velocity distribution of our experimental data along the x- and y-directions for discharge parameters of V$_d = 310$ volt and P = 14 Pascal. The distribution function are found to be maxwellian as can be seen from the fitted maxwellian function (solid line) on the experimental data points in the figures. The kinetic dust temperature is calculated from the width of the measured distribution by using the formula $E =\frac{1}{2}m \langle v^2_{x,y} \rangle = \frac{1}{2}k_B T_{x,y}$. 
Fig~\ref{fig:fig10} shows the measured kinetic temperature of the dust particles with background neutral pressure. It can be seen from the figure that at lower pressure the temperature is higher. It is important to be mentioned that our measured temperature is twice the order of magnitude less than previously reported temperature during melting transition in rf plasma. 
A rapid fall in the temperature is observed with increasing pressure and above 13 pascal it decays slowly and then saturates at $\sim$ 0.035 eV which is close to the room temperature. The decrease in temperature at increasing pressure can be directly attributed to the collisional cooling of the dust particles. We investigated mechanisms that could be responsible for the heating of the dust particle. Vaulina \textit{et al.} \cite{vaulina_charging} predicts the dust charge fluctuation is one of the main mechanisms to heat up the dust particles. The dust charge fluctuation is essentially depends on two parameters; particle charge and their mass. In order to estimate the dust charge, we have followed the procedure outlined in the paper by Khrapak \textit{et al.}\cite{khrapak_charging} that allows estimating the charge based on its dependence on the ratio of the charge density of the particles to that of the ions, defined as Havnes parameter \cite{havnes}  
\begin{equation}
P = \frac{a T_e}{e^2}\frac{n_p}{n_0}\cong 695 a T_e \frac{n_p}{n_0},    
\end{equation}
where a is the particle radius (in $\mu$m), T$_e$ is the electron temperature (in eV) and n$_p$ and n$_0$ is the particle and plasma densities, respectively. When P $>1$, the charge and floating potential are significantly diminished, while for P $\ll$ 1 the charge and floating potentials approach the values for an isolated particle \cite{melzer_book, vaulina_charging}. Knowing the information of P, one can estimate the dimensionless charge ($z = e^2 Z_d/4\pi\epsilon_0ak_BT_e$)\cite{khrapak_charging} and hence the corresponding charge residing on the particle, Q. For our range of pressure 15-11 Pascal and the corresponding measured parameters T$_e$ = (2.8-4.3) eV, n$_0$ = (1.5-0.8)$\times 10^{15}$/m$^3$, n$_p$ = (2.45-1.35)$\times10^{10}$/m$^3$, the dimensionless charge is estimated as z$\sim$2.25 and the value of charge varies from 1833-2984 (in the unit of e). This shows an increase in the charge at lower pressure. 
Therefore at lower pressure the charge fluctuation would be higher and hence particle get heated. This can lead to melting of the dust particle. In addition to the charge fluctuation, the streaming ions in also play a very important role in heating up the particles in dc glow discharge plasma. The ion streaming increases with decreasing pressure. However, the asymmetry of the electrodes where cathode tray are larger than the anode and the dust cloud are formed far away from the anode (not facing the anode)
has helped in reducing the heating effects associated with ion streaming. The ion path has been further influence by the metal confinement strips which helps ion streaming away from the region between the strips where dust crystal formed. A detailed investigation of such a behavior presented by Hari \textit{et al.}\cite{hari_pop_2018}. Thus with the help of present electrodes configuration and an additional confining strips
we are able to get an optimal condition of discharge parameter where we observed a stable crystal formation whereas a small variation in discharge parameters lead to melting unlike to rf plasma where ion streaming is not so strong and hence a large variation in discharge parameter breaks the crystalline structure into fluid-like state. 

\section{Conclusion \label{conclusion}}
In conclusion, we present an experimental investigation on the melting transition of dust crystal in DC glow discharge plasma. An asymmetrical electrode configuration along with the confining strips facilitates the formation of a hexagonal crystalline structure at a pressure of 15 Pascal and a transition from solid state to a fluid state is observed at reducing gas pressure. The melting transition of the crystal is demonstrated by investigating the global and local structural properties of the dust cloud as a function of neutral pressure. The pair correlation show a long - range order in the particle arrangement above 13 Pascal whereas ordering is reduced at lower pressure. The local variation in the structural dynamics of the particle cloud is examined by calculating the local bond order parameter $\Psi_6$ and defect fraction. The average bond order parameter is found to be maximum at 15 Pascal, $\Psi_6 = 0.87$ whereas it decreases to 0.4 at a pressure of 11 Pascal. The defect measurement does not show a strong correlation with the KTHNY theory of melting. The kinetic temperature is estimated from the velocity distribution of the particle. The measured kinetic dust temperature is found to be close to room temperature at higher pressure whereas it increases to 0.15 eV at lower pressure. We believe that a the combined effect of charge fluctuation and ion streaming yield an increased dust temperature which induced melting of the crystalline structure at lower pressure. Our findings could be useful for exploring other innovative modifications in various DC glow discharge devices to make them suitable for the study of dusty plasma crystals and related phenomena.  
\section{Acknowledgments}
Authors acknowledge the Institute for plasma research for providing the infrastructure. S. Jaiswal is thankful to Dr. S. k. Mishra for valuable discussions and S. K. Pandey and R. Mukherjee for technical help. Additional funding support was provided by the NSF EPSCoR program (OIA-1655280).
\section{References.\label{bibby}}

\end{document}